# Electric double-layer transistor using layered iron selenide Mott insulator TlFe$_{1.6}$Se$_2$

Takayoshi Katase,[1,4] Hidenori Hiramatsu,[2,3] Toshio Kamiya,[2,3] and Hideo Hosono[1,2,3,*]

[1] Frontier Research Center, Tokyo Institute of Technology, Mailbox S2-13, 4259 Nagatsuta-cho, Midori-ku, Yokohama 226-8503, Japan

[2] Materials and Structures Laboratory, Tokyo Institute of Technology, Mailbox R3-1, 4259 Nagatsuta-cho, Midori-ku, Yokohama 226-8503, Japan

[3] Materials Research Center for Element Strategy, Tokyo Institute of Technology, Mailbox S2-16, 4259 Nagatsuta-cho, Midori-ku, Yokohama 226-8503, Japan

[4] Present address: Research Institute for Electronic Science, Hokkaido University, Sapporo 001-0020, Japan

[*]Corresponding author. E-mail: hosono@msl.titech.ac.jp

**Abstract**

$A_{1-x}Fe_{2-y}Se_2$ ($A$ = K, Cs, Rb, Tl) are recently discovered iron-based superconductors with critical temperatures ($T_c$) ranging up to 32 K. Their parent phases have unique properties when compared with other iron-based superconductors; e.g., their crystal structures include ordered Fe vacancies, their normal states are antiferromagnetic (AFM) insulating phases, and they have extremely high Néel transition temperatures. However, control of carrier doping into the parent AFM insulators has been difficult due to their intrinsic phase separation. Here, we fabricated an Fe-vacancy-ordered $TlFe_{1.6}Se_2$ insulating epitaxial film with an atomically flat surface and examined its electrostatic carrier doping using an electric double-layer transistor (EDLT) structure with an ionic liquid gate. The positive gate voltage gave conductance modulation of three orders of magnitude at 25 K, and further induced and manipulated a phase transition; i.e., delocalized carrier generation by electrostatic doping is the origin of the phase transition. This is the first demonstration, to the authors' knowledge, of an EDLT using a Mott insulator iron-selenide channel and opens a way to explore high-$T_c$ superconductivity in iron-based layered materials, where carrier doping by conventional chemical means is difficult.

Layered copper-based oxides (cuprates) and iron-based superconductors are the most well-known types of superconductor because of their high $T_c$ of more than 50 K [1,2]. One common feature of these materials is that superconductivity emerges when the long-range AFM order in the parent phase is suppressed and varnished by carrier doping. However, the maximum $T_c$ of the cuprates (134 K for $HgBa_2Ca_2Cu_3O_{8+\delta}$) [3] is much higher than that of the iron-based materials (55 K for $SmFeAs(O_{1-x}F_x)$) [4]. This difference is related to the different electron correlation interactions of the materials; i.e., the parent phases of the cuprates are AFM Mott insulators, where the electron-electron Coulomb interaction is very strong, while the iron-based parent phases are AFM metals with weaker electron correlation. Related to this difference, the Néel transition temperatures ($T_N$) of the cuprate parent phases (e.g. 420 K for $YBa_2Cu_3O_6$) [5] are much higher than those of the iron-based phases (e.g. 140 K for SmFeAsO) [6]. Based on the high-$T_c$ cuprate scenario, it is considered that a key strategy to obtaining higher $T_c$ in the iron-based superconductors is to dope carriers into the parent phase of a Mott insulator with a higher $T_N$ and then disperse the magnetic order by carrier doping.

Recently, superconductivity was discovered at 32 K in a layered iron-selenide, $K_{0.8}Fe_2Se_2$ [7], which has drawn considerable attention, because this is the first material with a parent phase to exhibit an AFM insulating state among the iron-based superconductors [8]. The $A_{1-x}Fe_{2-y}Se_2$ ($A$ = K, Cs, Rb, Tl) system has the same crystal structure as that of the 122-type iron-based superconductor $BaFe_2As_2$ with a tetragonal $ThCr_2Si_2$-type structure [9] (**Fig. 1(a)**), and is composed of alternately stacked $A$ and FeSe layers along the $c$-axis. The Fe atoms in the FeSe layer form a square lattice, where the Se atoms are located at the apical sites of the edge-shared $FeSe_4$ tetrahedra.

However, the ideal chemical formula is $A_2Fe_4Se_5$ (the 245 phase), which satisfies charge neutrality conditions with consideration of formal ion charges (i.e., +1 for $A$, +2 for Fe, and −2 for Se); therefore, the $A$ and Fe sites include vacancies in the $ThCr_2Si_2$-type structure. It was shown that the Fe vacancies ($V_{Fe}$) in the parent 245 phase exhibit an order-disorder transition at ~500 K and form a $\sqrt{5} \times \sqrt{5} \times 1$ supercell (the unit cell formula is $A_8Fe_{16}Se_{20}$ with four Fe vacancies) (10), as shown in **Fig. 1**. Theoretical calculations suggested that the parent 245 phase is a Mott insulator with a Mott gap of ~100 meV (11,12). The gap was confirmed experimentally to be ~430 meV (13). The 245 Mott insulator exhibits an AFM long-range order with $T_N$ as high as 470−560 K, similar to that of the cuprates, along with an ordered magnetic moment of more than 3 Bohr magneton ($\mu_B$) at 10 K (10). Because of this similarity to the cuprates (i.e., a Mott insulator with high $T_N$), it is expected that a much higher $T_c$ would be realized in a $V_{Fe}$-ordered AFM insulator of 245 phase if the AFM order is suppressed by carrier doping.

Indeed, chemical electron-doping of the 245 phase has been performed by reducing the $V_{Fe}$ ($2–y > 1.93$), which induced superconductivity at ~30 K (8). However, it has been reported that these $A_{1-x}Fe_{2-y}Se_2$ superconductors intrinsically include phase separation into the superconducting phase, which is believed to exist in a phase without the $V_{Fe}$ order, and the AFM insulating phase ($V_{Fe}$-ordered 245 phase) (13). Further, it is controversy whether the superconducting phase is an intercalated phase like $Rb_{0.3}Fe_2Se_2$ (14), a $V_{Fe}$-free phase like $KFe_2Se_2$ (13), or this disordered $V_{Fe}$ phase (15). The coexistence of such multi-phases indicates that a well-controlled carrier doping structure of the 245 phase has yet to be realized by chemical-composition doping nor substitution, and the carrier doping effects are not yet clear. In contrast, carrier doping by an electrostatic method that uses a field-effect transistor structure is free from this structural alternation and would be suitable for study of phase transitions in $V_{Fe}$-ordered 245 Mott insulators that have never become a superconductor.

In this study, we focused on epitaxial films of one iron-selenide Mott insulator, TlFe$_{1.6}$Se$_2$, because TlFe$_{2-y}$Se$_2$ is much stable in air than the other $A_{1-x}$Fe$_{2-y}$Se$_2$ ($A$ = K, Rb, Cs) (16), and a fully $V_{Fe}$-ordered phase with high chemical homogeneity has been obtained in single crystals due to the lower vapour pressure of Tl than those of alkaline metals (17). In addition, the number of $V_{Fe}$ in TlFe$_{2-y}$Se$_2$ cannot be controlled over a wide range (the maximum 2–$y$ value is limited to only 1.6) regardless of the starting nominal compositions, and thus superconductivity has not previously been observed in TlFe$_{1.6}$Se$_2$ (16), although a bulk superconductivity was observed in mixed (Tl,K)$_{1-x}$Fe$_{2-y}$Se$_2$ (18). These features demonstrate that TlFe$_{1.6}$Se$_2$ AFM insulator is the most ideal target to examine electrostatic carrier doping. We therefore used an electric-double-layer transistor (EDLT) structure because the ionic liquid gate works as a nanometer-thick capacitor with a large capacitance and provides an effective way to accumulate a very high carrier density (maximum sheet carrier density of approximately $10^{15}$ cm$^{-2}$ under small gate voltages of around ±3 V). This high carrier modulation by the EDLT can alter electronic states over a very wide range and convert even a band insulator into a metal, and further into a superconductor (19); similarly, even a Mott insulator is converted into a metal (20). We therefore expected that the EDLT structure would also modulate the carrier density sufficiently to induce a phase transition such as superconductivity in the iron-selenide Mott insulator without chemical doping or structural alternation. Here, we used a single-phase (that means, homogeneous in structure, chemical composition and vacancy distribution) and $V_{Fe}$-ordered TlFe$_{1.6}$Se$_2$ insulating epitaxial film grown by pulsed laser deposition (PLD) with an atomically flat surface as the transport channel layer of the EDLT. Large field-effect current modulation was demonstrated in the EDLT, particularly at low temperatures. The electric field clearly decreased the activation energy and also induced a phase transition.

**Figure 2(a)** shows the out-of-plane X-ray diffraction (XRD) pattern of a $TlFe_{1.6}Se_2$ thin film grown at the optimum temperature of 600°C. Only the sharp peaks of the 00*l* diffractions of the $TlFe_{1.6}Se_2$ phase were observed, along with those of the $CaF_2$ substrate, indicating that the film grows along the [00*l*] direction. Although the in-plane lattice parameter of $CaF_2$ ($a/\sqrt{2}$ = 0.386 nm) is almost the same as that of $(La,Sr)(Al,Ta)O_3$ (LSAT, $a/2$ = 0.387 nm), the full width at half maximum (FWHM) values of the 004 rocking curve ($\Delta\omega$) of the film are much smaller when grown on the $CaF_2$ substrate (0.08°) than that grown on the LSAT substrate (0.8°) (**Figure 2(b)**). This suggests that an interface reaction occurs on the oxide LSAT substrate, while the fluoride $CaF_2$ substrate is more suitable for $TlFe_{1.6}Se_2$; similar results are reported also for iron-chalcogenide $FeSe_{0.5}Te_{0.5}$ epitaxial films (21). To confirm the epitaxial relationship between the $TlFe_{1.6}Se_2$ film and the $CaF_2$ substrate, asymmetric diffractions were measured. **Figure 2(c)** shows the results of $\phi$ scans of the 123 diffraction of the $TlFe_{1.6}Se_2$ film and the 202 diffraction of the $CaF_2$ substrate. Both peaks appear every 90° and exhibit four-fold symmetry, substantiating the heteroepitaxial nature of the $TlFe_{1.6}Se_2$ film growth on the $CaF_2$ substrate. Each peak (FWHM value = 0.2°) of the $TlFe_{1.6}Se_2$ film is rotated by 45° with respect to the peaks of the $CaF_2$ substrate, showing that the $TlFe_{1.6}Se_2$ film grows on the $CaF_2$ substrate with epitaxial relationships of [001] $TlFe_{1.6}Se_2$//[001] $CaF_2$ (out-of-plane) and [310] $TlFe_{1.6}Se_2$//[100] $CaF_2$ (in-plane). These epitaxial relationships are a natural consequence of the smallest in-plane lattice mismatching ($\Delta(d_{Tl-Tl} - d_{Ca-Ca})/d_{Ca-Ca} \times 100$ = 0.8%) as shown in **Fig.**

**2(d)**. **Figure 2(e)** shows the surface morphology of the $TlFe_{1.6}Se_2$ epitaxial film on the $CaF_2$ substrate. A flat surface with a step-and-terrace structure (root-mean-square roughness of 1.4 nm) was observed, indicating the layer-by-layer growth of $TlFe_{1.6}Se_2$ epitaxial films under optimized conditions. This result is also consistent with the observation of Pendellösung interference fringes (inset of **Fig. 2(a)**). The step height (**Fig. 2(f)**) observed by atomic force microscopy is ~0.7 nm, which agrees well with the distance between the nearest-neighbour FeSe–FeSe layers (corresponding to a half unit of the *c*-axis length (1.397 nm) of the $TlFe_{1.6}Se_2$ unit cell indicated in **Fig. 1(a)**). These results guarantee that the $TlFe_{1.6}Se_2$ epitaxial film has sufficiently high quality to be used for the EDLT transport channel.

The atomic structure and $V_{Fe}$ ordering in the $TlFe_{1.6}Se_2$ epitaxial film were examined by high-angle annular dark field scanning transmission electron microscopy (HAADF-STEM) and selected area electron diffraction (SAED). **Figures 3(a) and (b)** show the plan-view HAADF-STEM images of the $TlFe_{1.6}Se_2$ epitaxial film. $V_{Fe}$ are detected as the dark regions due to the enhanced Z-contrast of HAADF, showing the long-range periodic $V_{Fe}$ ordering. In addition, superlattice diffractions due to the $V_{Fe}$ ordering, similar to that of $TlFe_{1.6}Se_2$ single crystal (17), were observed in the SAED pattern **(c)** as indicated by $q_1$ and $q_2$. These results substantiate that the present sample is of a highly $V_{Fe}$-ordered phase. **Figure 3(d)** visualizes the arrangement of $V_{Fe}$ more clearly by the yellow lines superimposed on the HAADF-STEM image of **(a)**. Fully ordered $V_{Fe}$ are dominant in almost the whole region, while small phase separation to disordered-$V_{Fe}$ regions $\leq 5$ nm in size (the unmarked regions in **(d)**) were also observed by keeping the perfect coherency of the fundamental crystal structure.

A 20 nm-thick $TlFe_{1.6}Se_2$ epitaxial film was used as the EDLT transport channel.

**Figure 4** shows a schematic illustration of the EDLT, in which a six-terminal Hall bar channel and Au pad electrodes were formed using shadow masks. After pouring the ionic liquid, N,N-diethyl-N-methyl-N-(2-methoxyethyl)-ammonium bis-(trifluoromethylsulfonyl) imide (DEME-TFSI), into a silica glass cup, a Pt coil electrode was inserted into the ionic liquid to act as a gate electrode. The transfer curves (gate voltage ($V_G$) dependence of drain current ($I_D$)) at a drain voltage ($V_D$) of +0.3 V and output curves ($I_D$ vs. $V_D$ under various $V_G$) of the EDLT were then measured.

Figure 5(a) shows the cyclic transfer characteristics ($I_D$ vs. $V_G$) of the $TlFe_{1.6}Se_2$ EDLT at $T$=280 K. A positive $V_G$ of up to +4 V was applied to the Pt coil gate electrode, which accumulates electrons at the interface. When $V_G$ = +1.7 V was applied, $I_D$ began to increase. The maximum $I_D$ in the transfer curve reached 12 μA at $V_G$ = +4 V, along with a small on-off ratio of 1.5. The gate leakage current ($I_G$) (shown in the bottom panel of **Fig. 5(a)**) also increased at $V_G$ up to +4.0 V but was clearly smaller than $I_D$ in the whole $V_G$ region. After applying $V_G$ = +4 V, $I_D$ recovered to the initial values of 8 μA when $V_G$ was decreased to 0 V. The large hysteresis loop is observed due to the slow response of the ion displacement in the ion liquid. Probably due to the same reason, some parallel shift remains in the second $I_D$–$V_G$ loop; however, the shape and the hysteresis width are very similar to those of the first loop, guaranteeing the observed results are reversible and reproducible. These results demonstrate the electrostatic nature of carrier accumulation. In the output characteristics (Fig**. 5(b)**), the conductance $dI_D/dV_D$ increased with increasing $V_G$ at ≥ +2.0 V. Two output characteristics, which were measured before and after the transfer curve measurements in **(a)**, remain unchanged, which further guarantees the reversibility of the EDLT characteristics. However, the $I_D$ modulation is small at 280 K because of the high conductance at $V_G$ = 0, which originates from the highly naturally-doped carriers in the $TlFe_{1.6}Se_2$ film, as reported for a $TlFe_{1.6}Se_2$ bulk crystal in which the carrier density was estimated to be $\sim5\times10^{21}$ $cm^{-3}$ at $T$ = 150 K (16). Using the reported gate capacitance value of $\sim$10 $\mu F/cm^2$ (22), the maximum accumulated carrier density is estimated to be $2.5\times10^{14}$ $cm^{-2}$ at $V_G$ = 4V, and the field-effect mobility in the linear region of the output characteristics is estimated to be 0.18 $cm^2/(V \cdot s)$ at $V_G$ = +4 V. To estimate the carrier density induced in the $TlFe_{1.6}Se_2$ EDLT, we performed Hall effect measurements by applying magnetic fields of up to 9 T at temperatures between 300 and 25 K, but the Hall voltages ($V_{xy}$) obtained were below the detection limit of our measurement system. This suggests that the Hall mobility is smaller than 0.02 $cm^2/(V \cdot s)$, which is roughly consistent with the small field-effect mobility above. **Figure 5(c)** plots the $V_G$ dependence of the sheet conductance ($G_s$) at $T$ = 300–25 K. The $V_G$ dependences of $G_s$ were reversible also against repeated variation of measurement temperature (compare the open symbols and the closed symbols in **Fig. 6(a)**). With decreasing $T$, $G_s$ at $V_G$ = 0 V steeply decreased from $2.2\times10^{-5}$ to $1.5\times10^{-8}$ S because of the decrease in carrier density. It should be noted that large $G_s$ modulation with gains of three orders of magnitude was demonstrated at $T$ = 25 K.

**Figure 6(a)** shows the $T$ dependences of the sheet resistance $R_s$ ($R_s$–$T$) for the $TlFe_{1.6}Se_2$ EDLT at $V_G$ = 0, +2, and +4 V. The $R_s$–$T$ characteristics from 300 to 30 K at $V_G$ = 0 V indicate simple thermally-activated behaviour, given by $R_s = R_{s0}\exp(E_a/k_BT)$ (where $R_{s0}$ is a constant, $k_B$ is the Boltzmann constant, and $E_a$ is the activation energy). This trend is similar to the $R_s$–$T$ behaviour of insulating $TlFe_{2-y}Se_2$ single crystals with $2-y < 1.5$ (18), but the resistivity anomaly due to a magnetic phase transition of spin re-orientation at 100 K observed in $TlFe_{1.6}Se_2$ single crystal (16,17) was not detected.

The $E_a$ value of the TlFe$_{1.6}$Se$_2$ EDLT at $V_G = 0$ obtained from the Arrhenius fitting is 20 meV (**Fig. 6(b)**). This value is smaller than 57.7 meV of TlFe$_{1.47}$Se$_2$ single crystal (18) but almost double of that of TlFe$_{1.6}$Se$_2$ single-crystals (11 meV) (16). These observations indicate the naturally-doped carrier density should be smaller than that of the TlFe$_{1.6}$Se$_2$ single crystals. On the other hand, The $E_a$ is far smaller than the calculated (~100 meV) (11,12) and experimentally measured Mott gaps (~430 meV) (13), which suggests that the TlFe$_{1.6}$Se$_2$ film is doped with carriers.

The $R_s$–$T$ behaviour largely varied with $V_G$, particularly in the low-$T$ region, which is seen also in **Fig. 5(c)**. The $R_s$–$T$ curves were reversible in the cooling and heating cycles. The $E_a$ value estimated in the high-$T$ region decreased from ~20 meV at $V_G = 0$–2 V to 8.9 meV at $V_G = +4$ V (**Fig. 6(b)**); i.e., $R_s$ at $V_G = +4$ V is almost independent of $T$, indicating that a highly accumulated channel was formed by the application of $V_G$. The $R_s$–$T$ curves at $V_G \geq 2$ V do not show a simple thermally-activated behaviour and exhibit humps at $T = 55$ K for $V_G = +2$ V and at 40 K for $V_G = +4$ V (indicated by vertical arrows in **Fig. 6(a)**). That is, when $V_G$ was increased from 0 to +2 V, the resistance hump appeared at $T_{hump} = 55$ K, and $R_s$ increased steeply again at $T \leq 31$ K. At $V_G = +4$ V, $T_{hump}$ shifted to a lower $T$ (40 K), and $R_s$ levelled off in the further lower $T$ region. It would be possible to consider that the resistivity humps is attributed to a precursory phenomenon of a metal-insulator (MI) transition because $E_a$ decreased sharply as confirmed in **Fig. 6(b)**. As for $A_{1-x}$Fe$_{2-y}$Se$_2$ superconductor single crystals, they also exhibit resistance humps and crossovers from an insulating state to a metallic state at $T_{hump}$, and finally to a superconducting state (18). In the case of (Tl,K)Fe$_{2-y}$Se$_2$ in literature (inset to **Fig. 6(a)**) (18), the resistance hump appears at $2-y = 1.68$, and the superconductivity appears at $2-y \geq 1.76$. This value is 16 % larger than that of the TlFe$_{1.6}$Se$_2$ film in this work. These preceding works suggest that the $T_{hump}$ observed in this study can also be related to the MI transition and superconductivity; however, we could not observe superconductivity up to the maximum $V_G$ of +4 V in this study.

A similar phenomenon, which is attributed to a magnetic phase transition, has been observed also in fully $V_{Fe}$-ordered (100 K) (17) and multi-phase (100 – 150 K) (23) TlFe$_{1.6}$Se$_2$ single crystals. On the other hand, the resistance humps are attributed to the formation of an orbital-selective Mott phase (OSMP) for the $V_{Fe}$-poor $A_{1-x}$Fe$_{2-y}$Se$_2$ bulk crystals, where one of the Fe 3$d$ orbitals, $d_{xy}$, remains localized and the other four orbitals are delocalized (24,25). The Mott insulator phase dominates the resistance above $T_{hump}$, while the OSMP prevails below it; finally, superconductivity appears below $T_{hump}$ of the OSMP transition at higher carrier doping levels. Due to these similarities, we consider that the resistance humps observed in this study are more likely assigned to the same origin of magnetic phase transition or OSMP.

In summary, electrostatic carrier doping into the Mott insulator iron-selenide $V_{Fe}$-ordered TlFe$_{1.6}$Se$_2$ by the EDLT structure was demonstrated using a single-phase epitaxial film with an atomically flat surface grown on a CaF$_2$ substrate. The EDLT structure, based on an ionic-liquid gate, successfully controlled the conductance and induced the phase transition assignable to a magnetic phase transition or OSMP. This demonstration of carrier doping of the Mott insulator iron-selenide by the electrostatic method offers a way to extend the exploration of high-$T_c$ superconductors even to insulating materials, in which chemical doping methods do not work.

## Experiments

*Film growth and characterization of film:* Epitaxial films of TlFe$_{1.6}$Se$_2$ were grown on fluorite-type $CaF_2$ and mixed perovskite type $(La,Sr)(Al,Ta)O_3$ (LSAT) (001) single crystals by PLD. A KrF excimer laser (wavelength of 248 nm) was used to ablate a $TlFe_{1.6}Se_2$ polycrystalline target disk, which was synthesized using a two-step solid-state reaction. Fine pieces of the Tl metal and powders of FeSe and Se were mixed in a stoichiometric atomic ratio of Tl:FeSe:Se = 1:1.6:0.4 and sealed in an Ar-filled stainless-steel tube. The mixture was first reacted at 400 °C for 5 h, and then at 650 °C for 10 h. The resulting powders were ground thoroughly and pressed into pellets, which were then placed in Ar-filled stainless-steel tubes and heated at 650 °C for 16 h. All the PLD target fabrication procedures other than the heating process were carried out in an Ar-filled glove box. The base pressure of the PLD growth chamber used in this study is ~$10^{-5}$ Pa. The laser energy fluence and the repetition rate were 10 J/cm$^2$ and 10 Hz, respectively. When grown in the 300–550 °C temperature range, epitaxial films were obtained, but their surfaces were relatively rough because of the three-dimensional growth mode, while the FeSe impurity phase was detected at temperatures ≥650 °C. Thus, we concluded that the optimal growth temperature was 600 °C.

The film structures, including the crystalline quality and the orientation of the crystallites, were examined by XRD (anode radiation: monochromatic CuK$\alpha_1$). The film thickness was characterized by X-ray reflectivity. The chemical composition of the film was checked by X-ray fluorescence measurements and electron probe micro-analyser (EPMA), and confirmed that the film's chemical composition is the same as that of the PLD target. EPMA mapping indicated that composition of the epitaxial films were homogeneous with a spatial resolution of a few micrometers. The surface morphology of the film was measured with an atomic force microscope. The microstructure of $V_{Fe}$ ordering in $TlFe_{1.6}Se_2$ epitaxial films was examined by HAADF-STEM and SAED. The STEM sample was prepared with a focused ion beam system. All these characterization measurements were performed at room temperature.

*Device fabrication and electrical properties characterization:* The 20 nm-thick $TlFe_{1.6}Se_2$ epitaxial films on the $CaF_2$ (001) substrate were used as the transport channel of the EDLT. The $TlFe_{1.6}Se_2$ channel layer with a six-terminal Hall bar geometry (channel size: 500 μm long and 200 μm wide) and the Au pad electrodes were deposited using shadow masks. After bonding Au wires to the Au pads with In metal, a silica glass cup was placed on the devices and the Au wires were fixed with an epoxy adhesive. We used an ionic liquid, DEME-TFSI, as the medium for the gate electrode because it has a wide electrochemical potential window that extends up to +4 V and it is free from water, which means that it is suitable for application to the EDLT. The ionic liquid was used to fill the silica-glass cup and then a Pt coil was inserted into the ionic liquid to act as the gate electrode.

Transfer curves (i.e., the $V_G$ dependence of $I_D$) and the output curves ($I_D$ vs. $V_D$ under various values of $V_G$) were taken from the results gathered by a source measurement unit. The temperature ($T$) dependence of $R_s$ was measured by the four-probe method over a $T$ range of 2–300 K. Because DEME-TFSI exhibits a glass transition from a rubber phase to a glass phase at $T$ = 190 K, and the ion motion is frozen out at lower values of $T$, $V_G$ was applied at 300 K to form a highly-accumulated gate structure, and then $T$ was reduced while maintaining the same $V_G$ (26).

# References

1. Bednorz J G and Müller K A (1986) Possible high $T_c$ superconductivity in the Ba-La-Cu-O system. *Z Phys B* 64(2):189–193.

2. Kamihara Y, Watanabe T, Hirano M, Hosono H. (2008) Iron-based layered superconductor La[$O_{1-x}F_x$]FeAs ($x$ = 0.05–0.12) with $T_c$ = 26 K. *J Am Chem Soc* 130(11):3296 – 3297.

3. Schilling A, Cantoni M, Guo J D, Ott H R (1993) Superconductivity above 130 K in the Hg-Ba-Ca-Cu-O system. Nature 363(6424):56–58.

4. Ren Z A, Ju W, Yang J, Yi W, Shen X L, Li Z C, Che G C, Dong X L, Sun L L, Zhou F, Zhao Z X (2008) Superconductivity at 55 K in iron-based F-doped layered quaternary compound Sm[$O_{1-x}F_x$]FeAs. *Chin Phys Lett* 25(6):2215–2216.

5. Tranquada J M, Moudden A H, Goldman A I, Zolliker P, Cox D E, Shirane G, Sinha S K, Vaknin D, Johnston D C, Alvarez M S, Jacobson A J, Lewandowski J T, Newsam J M (1988) Antiferromagnetism in $YBa_2Cu_3O_{6+x}$. *Phys Rev B* 38(4):2477–2485.

6. Drew A J, Niedermayer Ch, Baker P J, Pratt F L, Blundell S J, Lancaster T, Liu R H, Wu G, Chen X H, Watanabe I, Malik V K, Dubroka A, Rossle M, Kim K W, Baines C, Bernhard C (2009) Coexistence of static magnetism and superconductivity in $SmFeAsO_{1-x}F_x$ as revealed by muon spin rotation. *Nat Mater* 8(4):310–314.

7. Guo J, Jin S, Wang G, Wang S, Zhu K, Zhou T, He M, Chen X (2010) Superconductivity in the iron selenide $K_xFe_2Se_2$ ($0{\leq}x{\leq}1.0$). *Phys Rev B* 82(18):180520.

8. Yan Y J, Zhang M, Wang A F, Ying J J, Li Z Y, Qin W, Luo X G, Li J Q, Hu J, Chen X H (2012) Electronic and magnetic phase diagram in $K_xFe_{2-y}Se_2$ superconductors. *Sci Rep* 2:212.

9. Rotter M, Tegel M, Johrendt D (2008) Superconductivity at 38 K in the iron arsenide ($Ba_{1-x}K_x$)$Fe_2As_2$. *Phys Rev Lett* 101(10):107006.

10. Ye F, Chi S, Bao W, Wang X F, Ying J J, Chen X H, Wang H D, Dong C H, Fang M (2011) Common crystalline and magnetic structure of superconducting $A_2Fe_4Se_5$ (*A*=K, Rb, Cs, Tl) single crystals measured using neutron diffraction. *Phys Rev Lett* 107(13):137003.

11. Yu R, Zhu J X, Si Q (2011) Mott transition in modulated lattices and parent insulator of $(K,Tl)_y Fe_x Se_2$ superconductors. *Phys Rev Lett* 106(18):186401.

12. Yan X W, Gao M, Lu Z Y, Xiang T (2011) Electronic structures and magnetic order of ordered-Fe-vacancy ternary iron selenides $TlFe_{1.5}Se_2$ and $AFe_{1.5}Se_2$ ($A$=K, Rb, or Cs). *Phys Rev Lett* 106(8):087005.

13. Li W, Ding H, Deng P, Chang K, Song C, He K, Wang L, Ma X, Hu J P, Chen X, Xue Q K (2012) Phase separation and magnetic order in K-doped iron selenide superconductor. *Nat Phys* 8(2):126–130.

14. Texier Y, Deisenhofer J, Tsurkan V, Loidl A, Inosov D S, Friemel G, Bobroff J (2012) NMR study in the iron-selenide $Rb_{0.74}Fe_{1.6}Se_2$: determination of the superconducting phase as iron vacancy-free $Rb_{0.3}Fe_2Se_2$. *Phys Rev Lett* 108(23):237002.

15. Chen F, Xu M, Ge Q Q, Zhang Y, Ye Z R, Yang L X, Jiang Juan, Xie B P, Che R C, Zhang M, Wang A F, Chen X H, Shen D W, Hu J P, Feng D L (2011) Electronic identification of the parental phases and mesoscopic phase separation of $K_x Fe_{2-y}Se_2$ superconductors. *Phys Rev X* 1(2):021020.

16. Sales B C, McGuire M A, May A F, Cao H, Chakoumakos B C, Sefat A S (2011) Unusual phase transitions and magnetoelastic coupling in $TlFe_{1.6}Se_2$ single crystals. *Phys Rev B* 83(22):224510.

17. May A F, McGuire M A, Cao H, Sergueev I, Cantoni C, Chakoumakos B C, Parker D S, Sales B C (2012) Spin reorientation in $TlFe_{1.6}Se_2$ with complete vacancy ordering. *Phys Rev Lett* 109(7):077003.

18. Fang M H, Wang H D, Dong C H, Li Z J, Feng C M, Chen J, Yuan H Q (2011) Fe-based superconductivity with $T_c$=31 K bordering an antiferromagnetic insulator in $(Tl,K) Fe_x Se_2$. *Europhys Lett* 94(2):27009.

19. Ueno K, Nakamura S, Shomotani H, Ohtomo A, Kimura N, Nojima T, Aoki H, Iwasa Y, Kawasaki M (2008) Electric-field-induced superconductivity in an insulator. *Nat Mater* 7(11): 855–858.

20. Nakano M, Shibuya K, Okuyama D, Hatano T, Ono S, Kawasaki M, Iwasa Y, Tokura Y (2012) Collective bulk carrier delocalization driven by electrostatic surface charge accumulation. *Nature* 487(7408):459–462.

21. Tsukada I, Hanawa M, Akiike T, Nabeshima F, Imai Y, Ichinose A, Komiya S, Hikage T, Kawaguchi T, Ikuta H, Maeda A (2011) Epitaxial growth of $FeSe_{0.5}Te_{0.5}$ thin films on $CaF_2$ substrates with high critical current density. *Appl Phys Express* 4(5):053101.

22. Ye J T, Inoue S, Kobayashi K, Kasahara Y, Yuan H T, Shimotani H, Iwasa Y (2010) Liquid-gated interface superconductivity on an atomically flat film. *Nat Mater* 9(2):125.

23. Cao H, Cantoni C, May A F, McGuire M A Chakoumakos B C, Pennycook S J, Custelcean R, Sefat A S, Sales B C (2012) Evolution of the nuclear and magnetic structures of $TlFe_{1.6}Se_2$ with temperature. *Phys Rev B* 85(5):054515.

24. Yi M, Lu D H, Yu R, Riggs S C, Chu J H, Lv B, Liu Z K, Lu M, Cui Y T, Hashimoto M, No S K, Hussain Z, Chu C W, Fisher I R, Si Q, Z.-X. Chen (2013) Observation of temperature-induced crossover to an orbital-selective Mott phase in $A_xFe_{2-y}Se_2$ ($A$=K, Rb) superconductors. *Phys Rev Lett* 110(6):067003.

25. Yu R, Si Q (2013) Orbital-selective Mott phase in multiorbital models for alkaline iron selenides $K_{1-x}Fe_{2-y}Se_2$. *Phys Rev Lett* 110(14):146402.

26. Yuan H T, Shimotani H, Tsukazaki A, Ohtomo A, Kawasaki M, Iwasa Y (2009) High-density carrier accumulation in ZnO field-effect transistors gated by electric double layers of ionic liquids. *Adv Funct Mater* 19(7):1046–1053.

**Acknowledgments**

This work was supported by the Japan Society for the Promotion of Science (JSPS), Japan, through the "Funding Program for World-Leading Innovative R&D on Science and Technology (FIRST Program)" and by Ministry of Education, Culture, Sports, Science and Technology (MEXT), Japan, through the "Element Strategy Initiative to Form Core Research Center."

## Figures

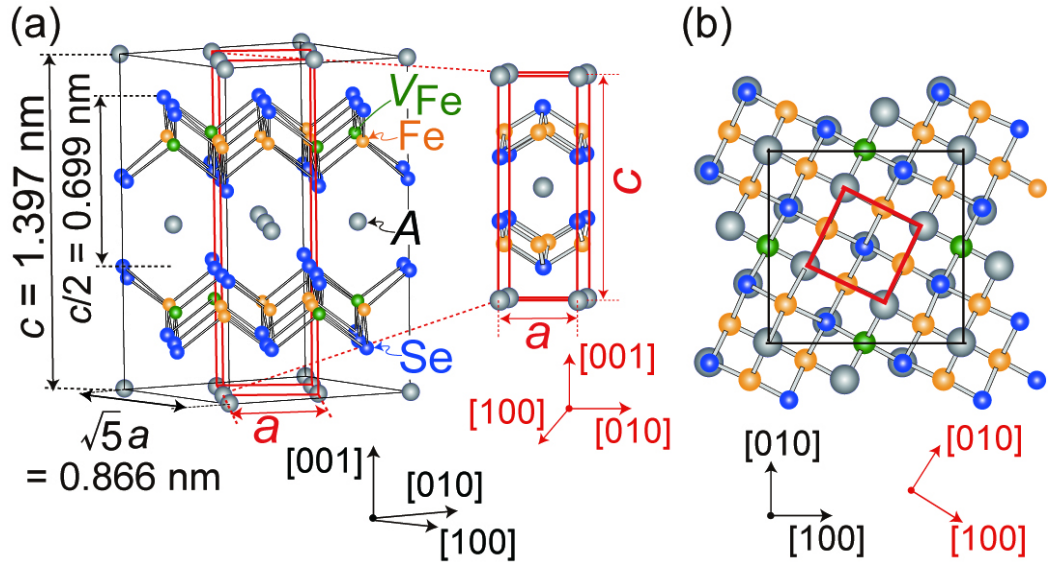

Fig. 1. Crystal structures of 122-type $A_{1-x}Fe_{2-y}Se_2$ ($A$ = K, Cs, Rb, Tl) and 245-type parent-phase $A_2Fe_4Se_5$ viewed along (a) the [120] and (b) the [001] directions. The spheres represent $A$ (gray), Fe (orange), Se (blue), and Fe vacancy sites ($V_{Fe}$, green). The 122-type tetragonal fundamental cell and the 245-type $\sqrt{5} \times \sqrt{5} \times 1$ supercell are indicated by the red and black lines, respectively.

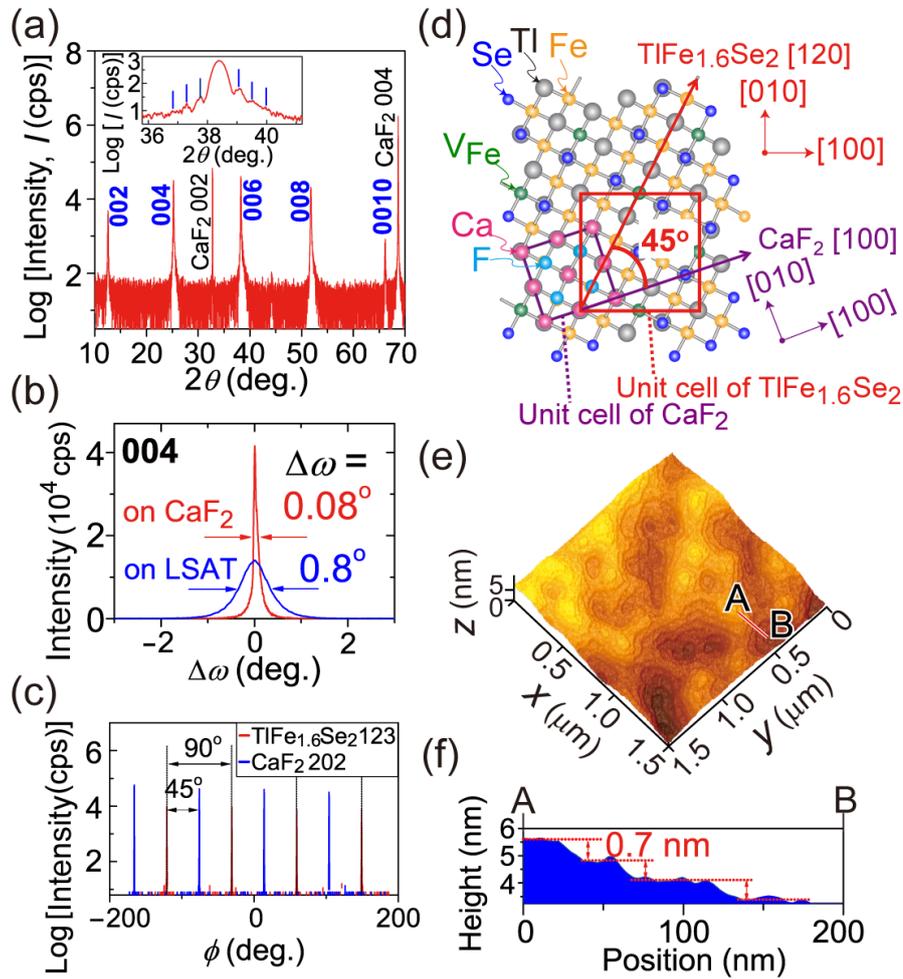

Fig. 2. (a) Out-of-plane XRD pattern of a TlFe$_{1.6}$Se$_2$ film grown on a CaF$_2$ (001) substrate. The inset shows the magnified pattern around the 006 diffraction. The vertical lines indicate the positions of the Pendellösung interference fringes. (b) Rocking curves of the 004 diffractions of the TlFe$_{1.6}$Se$_2$ films on CaF$_2$ and (La,Sr)(Al,Ta)O$_3$ (LSAT) substrates. (c) $\phi$ scans of the 123 diffraction of the TlFe$_{1.6}$Se$_2$ film and the 202 diffraction of the CaF$_2$ substrate. (d) In-plane atomic configuration of the TlFe$_{1.6}$Se$_2$ epitaxial film on the CaF$_2$ substrate. (e) Topographic atomic force microscopy image of the surface of the TlFe$_{1.6}$Se$_2$ epitaxial film on the CaF$_2$ substrate. (f) Height profile across the line A–B shown in (e).

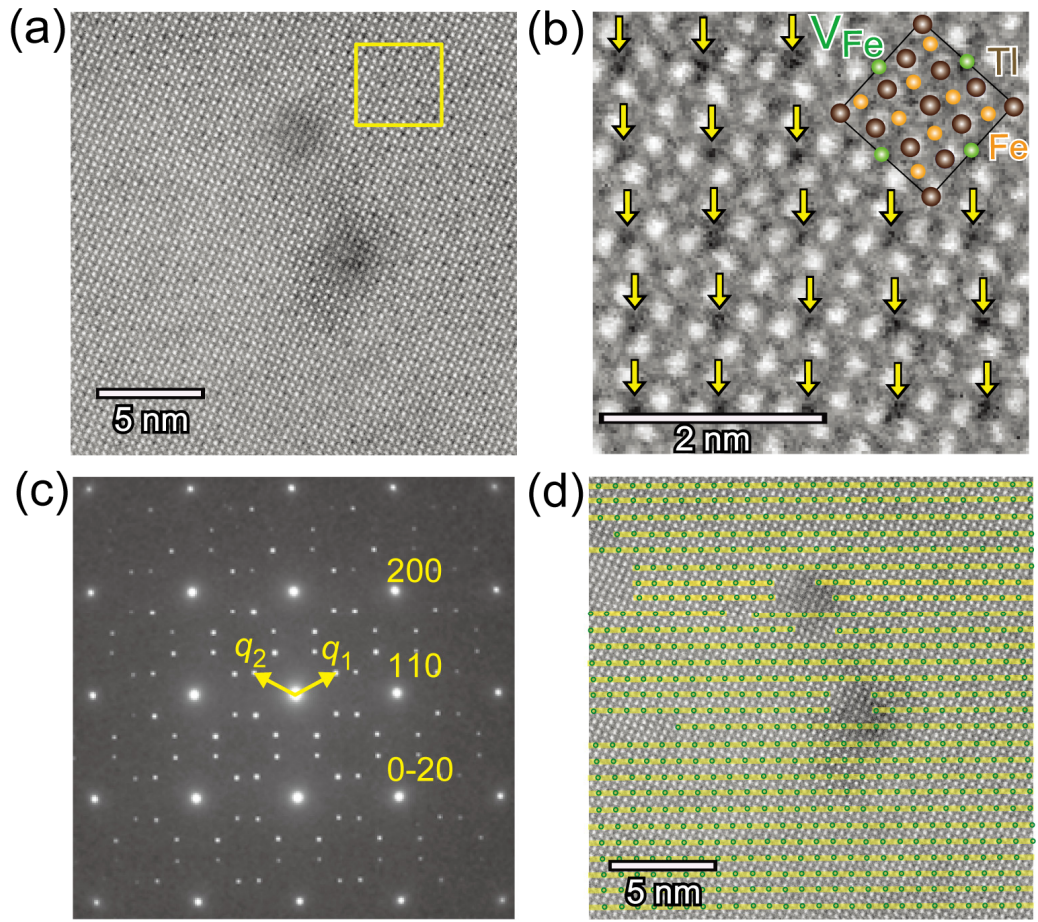

Fig. 3. (a) [001] plan-view HAADF-STEM image of TlFe$_{1.6}$Se$_2$ epitaxial film. (b) Magnified HAADF-STEM image of the yellow square region in (a). The vertical yellow arrows indicate the $V_{Fe}$ sites with dark contrast. Inset illustration shows the crystal structure of TlFe$_{1.6}$Se$_2$, where only Tl and Fe sites are shown because the positions of Se and Tl sites overlap over them (see Fig. 1(b)). The square shows the superlattice unit cell, where $V_{Fe}$ are shown by the green circles. (c) The SAED pattern with electron beam along [001]. Two superlattice reciprocal vectors due to $V_{Fe}$ ordering are indexed by $q_1$ and $q_2$. (d) The small green circles indicate all of detected $V_{Fe}$, and the arrangement of $V_{Fe}$ are indicated by yellow lines. The horizontal bar in each figure indicates the length scale.

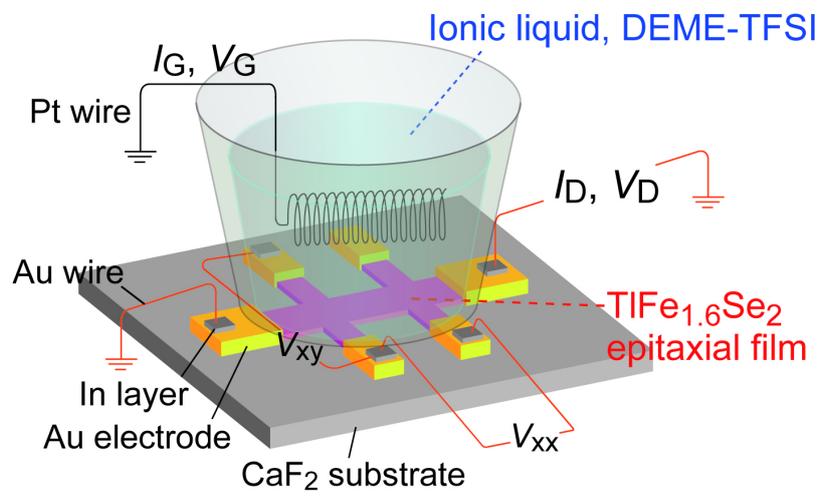

Fig. 4. Schematic image of the electric-double-layer transistor (EDLT) using the TlFe$_{1.6}$Se$_2$ epitaxial film with a six-terminal Hall bar structure on a CaF$_2$ substrate. $V_G$ was applied via a Pt counter-electrode through the ionic liquid, DEME-TFSI, contained in a silica-glass cup. Electrical contacts were formed using Au wires and In/Au metal pads.

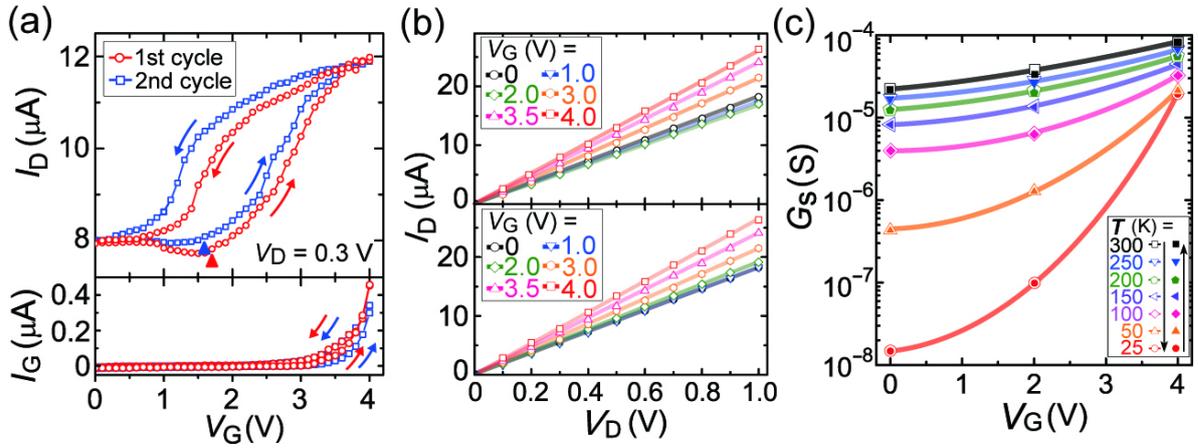

Fig. 5. (a) Transfer characteristics ($I_D$–$V_G$) at $V_D$ = +0.3 V and $T$ = 280 K cyclically measured for two loops. The arrows indicate the $V_G$-sweep directions, and the triangles show the positions where $I_D$ begins to increase. The leakage current ($I_G$) versus $V_G$ is also shown at the bottom. (b) Output characteristics ($I_D$–$V_D$) at $V_G$ = + 0–4 V and $T$ = 280 K, measured before (upper panel) and after (bottom panel) the transfer characteristics measurement in (a). (c) $V_G$ dependence of sheet conductance ($G_s$) measured with decreasing $T$ (open symbols) and increasing $T$ (closed symbols) over the 25–300 K range. The solid lines are guide for eyes to see the change in $G_s$ clearly.

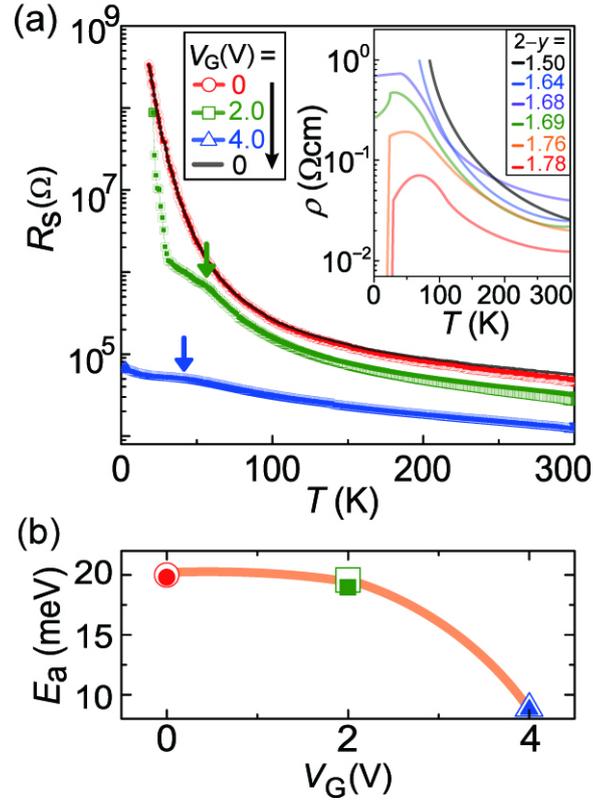

Fig. 6. (a) $T$ dependences of $R_s$ for the TlFe$_{1.6}$Se$_2$ EDLT measured with increasing $T$ (open symbols) and decreasing $T$ (closed symbols) at $V_G$ = 0→+2.0→+4.0→0 V. The arrows indicate the positions of resistance humps. The reported $\rho$–$T$ curves of (Tl,K)Fe$_{2-y}$Se$_2$ bulk materials (Ref. 18) are shown for comparison in the inset. A resistance hump appears at 2–$y$ ≥ 1.68, and superconductivity emerges at 2–$y$ ≥ 1.76. (b) The activation energy ($E_a$) estimated from (a) in the high-$T$ region as a function of $V_G$.